\pdfoutput=1
\documentclass[twocolumn,english,aps,superscriptaddress,prx]{revtex4-1}

\usepackage{babel}
\usepackage{amsmath}
\usepackage{amssymb}
\usepackage{graphicx}
\usepackage{lineno}

\newcommand{\appropto}{\mathrel{\vcenter{
  \offinterlineskip\halign{\hfil$##$\cr
    \propto\cr\noalign{\kern2pt}\sim\cr\noalign{\kern-2pt}}}}}

\begin{document}

\title{Critical properties of the Majorana chain with competing interactions}

\author{Natalia Chepiga}
\affiliation{Kavli Institute of Nanoscience, Delft University of Technology, Lorentzweg 1, 2628 CJ Delft, The Netherlands}

\date{\today}
\begin{abstract}
We explore critical properties of a chain of interacting Majorana fermions - particles that are their own antiparticles. We study the combined effect of two competing interaction terms of the shortest possible range and show this results in a very rich phase diagram with nine different phases five of which are critical. In addition, we report a wide variety of quantum phase transitions: the tri-critical Ising lines;  the Lifshitz critical line characterized by the dynamical critical exponent $z=3$; two Kosterlitz-Thouless transitions, and an exotic first order transition between the floating and the gapped phases. However, the most surprising result is the emergence of the commensurate line at which the floating phases collapse into direct transition. We provide numerical evidences that the resulting multicritical point belongs to the universality class of the eight-vertex model. Implications in the context of supersymmetric properties of the Majorana chain is briefly discussed.
\end{abstract}
\pacs{
75.10.Jm,75.10.Pq,75.40.Mg
}

\maketitle


\section{Introduction}

Over the years quantum critical phenomena are attracting a lot of interest of both theorists and experimentalists in condensed matter physics\cite{giamarchi,tsvelik}. The concept of universality classes and the invention of density matrix renormalization group (DMRG) algorithm\cite{dmrg1,dmrg2,dmrg3,dmrg4} open a possibility to study quantum phase transitions on simple lattice models and position strongly-correlated 1D systems as a fruitful playground to explore quantum critical phenomena.  Among various models of frustrated spin chains, bosons and fermions, a special role is played by the models of Majorana zero modes bridging various areas of physics. A model of non-interacting Majorana fermions - particles that are their own antiparticles - is rigorously equivalent to the model of hard-core bosons and spinless fermions with an extra term that simultaneously create or destroy a pair of particles. Revising this model as an effective model of $p$-wave superconductor Kitaev\cite{Kitaev_2001} has shown that it possesses Majorana edge states. 
 Motivated by their potential usage for qubits\cite{RevModPhys.80.1083,2015arXiv150102813D} this discovery launched a tremendous experimental activity\cite{majorana_exp1,2012NatPh...8..795R,Deng_2012,Das_2012,alicea,beenakker,Toskovic_2016}.

At the same time, and quite logically since fermions experience a repulsion,  theoretical studies of Majorana chains are centered  on extended models incorporating interactions between Majorana fermions in various forms\cite{PhysRevB.84.085114,PhysRevLett.115.166401,rahmani,PhysRevB.81.134509,PhysRevB.83.075103,katsura,ObrienFendley, verresen,chepiga8vertex,chepiga_laflorencie}.  In this paper we study the combined effect of the two interaction terms of the shortest possible range. The motivation behind that is twofold. Firstly, any realistic interaction potential is a continuous function that does not vanish immediately beyond the first- or second neighbors. Thus by considering the terms beyond the shortest possible range of interactions we effectively include the next-to-leading order corrections into the lattice Hamiltonian. Secondly,
competing interactions introduce frustration and may lead to a new critical behavior and exotic phenomena.
 The microscopic model can be defined by the following Hamiltonian:
   \begin{multline}
{\cal{H}}=\mathrm{i}t\sum_a \gamma_a\gamma_{a+1} -g\sum_a \gamma_a\gamma_{a+1}\gamma_{a+2}\gamma_{a+3}  \\-f\sum_a \gamma_a\gamma_{a+1}\gamma_{a+3}\gamma_{a+4},
\label{eq:majorana}
\end{multline}
where Majorana operators $\gamma_a$ are Hermitian and obey $\{\gamma_a,\gamma_{a^\prime} \}=2\delta_{a,a^\prime}$. Since $\gamma_a^2=1$, the shortest range non-trivial interaction $g$ spans over four consecutive sites. The last term $f$ spans the four-body Majorana operator over five consecutive sites. The precise form of this interaction term has been introduced by O'Brian and Fendley\cite{ObrienFendley} as a shortcut to realize an exact supersymmetry of the tri-critical Ising point on a lattice\cite{ObrienFendley}.

By means of Jordan-Wigner transformation, the Hamiltonian (\ref{eq:majorana}) can be written in terms of Pauli matrices $\sigma^{x,z}_j$:
\begin{multline}
  {\cal{H}}=\sum_j\left[-J\sigma^x_j\sigma^x_{j+1}-h\sigma^z_j+ g(\sigma^z_j\sigma^z_{j+1}+ \sigma^x_j\sigma^x_{j+2})\right.\\\left.+f(\sigma^z_{j}\sigma^x_{j+1}\sigma^x_{j+2}+\sigma^x_{j}\sigma^x_{j+1}\sigma^z_{j+2})\right].
  \label{eq:spin}
\end{multline}
Without loss of generality we set $J=1$ throughout the paper. The equivalence between two models is exact (up to boundary terms) when $J=h=t$.   In the first two terms one can immediately recognize the celebrated transverse-field Ising model\cite{lieb_two_1961,pfeuty}. It is therefore not surprising that in the non-interacting case $g=f=0$ the system is critical and belongs to the Ising universality class\cite{difrancesco}. The model with $g=0$ has been carefully studied by O'Brien and Fendley\cite{ObrienFendley}. It has been shown that there is a frustration-free point located at $f=1/2$  with three-fold degenerate ground-states: one $\mathbb{Z}_2$-symmetry preserving ground state coexisting with two symmetry-broken ones. Building on the intuition from the transverse-field Ising model, the entire gapped three-fold degenerate phase that hosts the frustration-free point corresponds to the first order phase transition that system undergoes upon tuning the transverse field $h$ from the two-fold degenerate $\mathbb{Z}_2$ phase to a paramagnetic one.
 The transition changes from Ising to first order at $f\approx0.428$\cite{ObrienFendley}. At the end point the critical behavior is described by the tri-critical Ising superconformal field theory\cite{ObrienFendley,difrancesco}. Similar to other frustration-free points, such as Majumdar-Ghosh point\cite{MajumdarGhosh} in spin-1/2 zig-zag chain and Affleck-Kennedy-Lieb-Tasaki point\cite{AKLT,PhysRevLett.77.5142} of the bilinear-biquadratic spin-1 chain, the  exact point of the Majorana chain at $f=1/2$ is also a disorder point, beyond which the system develops incommensurate short-range correlations.

The Majorana chain with $h=t=1$ and $g$ interaction only has been intensely studied in recent years\cite{PhysRevLett.115.166401,rahmani,PhysRevB.92.085139,chepiga_laflorencie}. It was shown that for small coupling $g$ the system remains in the critical Ising phase, however at $g\approx0.29$ it undergoes a Lifshitz transition into a critical phase with the central charge $c=3/2$\cite{rahmani,chepiga_laflorencie}. This phase is the Ising critical phase ($c=1/2$) superposed with the floating phase - Luttinger liquid phase ($c=1$) with incommensurate correlations. How and where this phase ends is a debated problem. 
According to Ref.\onlinecite{rahmani} and  Ref.\onlinecite{PhysRevB.92.085139} the floating phase, the Ising criticality and the incommensurability all terminate at the same point, though there was no agreement on the location of this transition: $g\approx 2.86$\cite{rahmani} vs $g\approx5$\cite{PhysRevB.92.085139}.
 Beyond this terminal point the system was expected to be gapped with four-fold degenerate ground-state\cite{{PhysRevB.92.085139},rahmani}.  However, in the recent study of an extended phase diagram with $J\neq h$\cite{chepiga_laflorencie} it was shown that the floating phase terminates much earlier - at $g\approx 1.3$, the Ising criticality terminates at $g\approx 3$  with the tri-critical Ising point, beyond which the system is indeed gapped, but the ground-state is six-fold degenerate.  It was also shown that incommensurability persists beyond $g=3$. Furthermore, the $g$-interaction also leads to an emergent supersymmetry: first, at the tri-critical Ising end point that was overlooked in early studies; second,  in the region where Luttinger liquid phase superposed with Ising critical line; and third, at the Kosterlitz-Thouless transition where the Ising critical line enters the floating phase\cite{huijse_mult,chepiga_laflorencie}. 
 
In this paper we study the combined effect of the two interaction terms introduced above. We show that: (i) the ground-state phase diagram is very rich and contains nine different phases, five of which are critical; (ii) there is a commensurate line along which two floating phases collapse into a single multicritical point in the universality class of the eight-vertex model; (iii) there is an extended Lifshitz critical line characterized by the dynamical critical exponent $z=3$; (iv) and finally, there are two critical lines effectively described by the tri-critical Ising superconformal field theory and at least one (and probably both) of them ends at the end point of the Lifshitz line. In addition, we argue that there might be an extended phase characterized by an emergent supersymmetry. 

We address the problem  numerically with state-of-the art DMRG\cite{dmrg1,dmrg2,dmrg3,dmrg4,dmrg_chepiga} algorithm written in terms of matrix product states. We perform simulations on a chains with up to $N=1201$ sites and with open and appropriately fixed boundary conditions. We perform up to 8 sweeps, keeping up to 3000 states and discarding the singular values below $10^{-8}$.

 The rest of the paper is organized as follows. In Section \ref{sec:pd} we overview the main features of the obtained phase diagram and discuss a self-duality of the model. In Section \ref{sec:TCI1} we discuss the tri-critical Ising and disorder lines. In Sec.\ref{sec:Lifshitz} we provide numerical evidences of an extended Lifhsitz critical line with the dynamical critical exponent $z=3$. In Sec.\ref{sec:floating} we provide the details of the three floating phases and discuss the Kosterlitz-Thouless transitions out of them. In Sec.\ref{sec:thetpoint} we provide numerical evidences that there is a commensurate line along which the two of the floating phases collapse into a single multicritical point. We also show that this point belongs to the universality class of the eight-vertex model.  Finally, we discuss the results and put them into a perspective in Sec.\ref{sec:discussion}.

\section{Phase diagram}
\label{sec:pd}

\begin{figure}[t!]
\centering 
\includegraphics[width=0.49\textwidth]{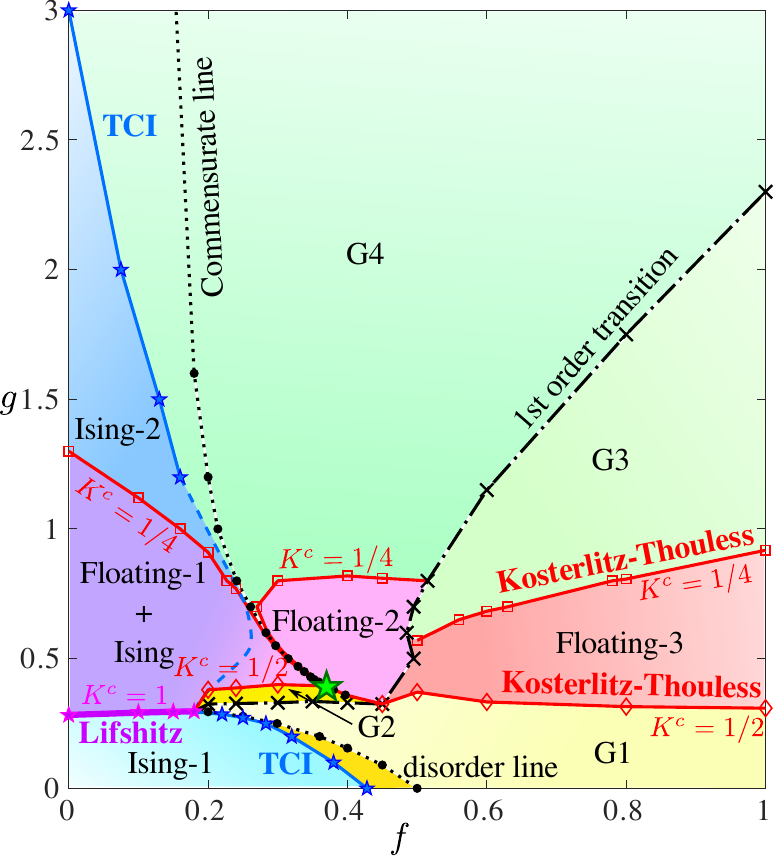}
\caption{{\bf Phase diagram of the interacting Majorana chain model Eq.(\ref{eq:spin}) as a function of coupling constants $g$ and $f$.} It contains four gapped phases: G1 (bright yellow) and G2 (pale and dark yellow) with three-fold degenerate ground-stats and G3 (pale green) and G4 (green) with six-fold degenerate ground-states. In addition, there are two critical phases in the Ising universality class (light and dark blue), two floating phases (magenta and rose); one critical phase where the floating phase is superposed with the Ising criticality (violet). The disorder line (dotted black) separates the commensurate region  (dark yellow) of the G1 phase from an incommensurate  one (pale yellow). Purple line denotes the Lifshitz critical line with $z=3$; solid blue lines state for the tri-critical Ising transitions; dashed blue lines are guide to eyes and indicate possible location of the tri-critical Ising line. Red lines indicate Kosterlitz-Thouless transitions taking place when the Luttinger liquid exponent $K$ reaches its critical values $1/4$ (red squares) and $1/2$ (red diamonds). Black dash-dotted line denotes the first order transition. Along the commensurate (dotted black) line the floating phases collapse into a multicritical point (green star) in the eight-vertex universality class.}
\label{fig:PD}
\end{figure}

Our main results are summarized in the  phase diagram presented in Fig.\ref{fig:PD}. It contains nine different phases  and various types of quantum phase transitions. Below we provide a short summary of different regimes.

{\bf Phases:}
\begin{itemize}
  \item {\bf Ising-1} is a critical phase realized for small couplings $f$ and $g$ that corresponds to the continuous phase transition in the Ising universality class between the topological $\mathbb{Z}_2$ phase for $h<1$ and a paramagnetic phase for $h>1$. This phase is commensurate.
  \item {\bf G1} is a gapped phase with three-fold degenerate ground-state that corresponds to the first order transition between the $\mathbb{Z}_2$ phase for $h<1$ and a paramagnetic phase for $h>1$. Small portion of the phase - before the disorder line - is commensurate; the rest of it has incommensurate short-range correlations.
  \item {\bf G2} is another gapped phase with three-fold degenerate ground-state. In contrast to G1 the states with spontaneously broken parity symmetry are the ground states for $h>1$.
  \item {\bf Floating-1 + Ising} is a critical phase where incommensurate Luttinger liquid (LL) is superposed with Ising criticality. The phase can be viewed as a continuous transition in the Ising universality class between two floating phases one of which (at $h>1$) spontaneously breaks $\mathbb{Z}_2$ symmetry. The entire phase is characterized by the central charge $c=3/2$ and extends from $f=0$ and almost to a commensurate line except a very tip of the phase where we see a clear indication that Ising criticality is no longer present. 
  \item {\bf Floating-2} phase corresponds to the fist order transition between the two floating phases with broken $\mathbb{Z}_2$ symmetry for $h>1$ and the one that preserves this symmetry for $h<1$. Due to the self-duality of the model the two floating phases are characterized by the same wave-vectors and same critical exponents.
  \item {\bf Floating-3} phase is a first order transition between two floating phases, but in contrast with Floating-2 the $\mathbb{Z}_2$ broken symmetry phase is located at $h<1$.
  \item {\bf Ising-2} is the Ising critical phase that for small $f$ extends beyond the Floating-1 phase. It corresponds to a continuous Ising transition between the period-2 phase for $h<1$ with spontaneously broken translation symmetry and period-2-$\mathbb{Z}_2$ phase for $h>1$ with both, translation and parity, symmetries broken.
  \item{\bf G4} is a gapped phase with six-fold degenerate ground-state that corresponds to the first order transition between the period-2 (at $h<1$) and period-2-$\mathbb{Z}_2$ (at $h>1$) phases.
  \item{\bf G3} is a gapped phase with six-fold degenerate ground-state that corresponds to the first order transition, but in contrast to the G4 phase the $\mathbb{Z}_2$ broken symmetry phase is realized at $h<1$.
\end{itemize}

In addition, the phase diagram contains a wide variety of quantum phase transitions, special lines and multicritical points.

{\bf Phase transitions and special lines:}
\begin{itemize}
  \item {\bf Tri-critical Ising (TCI):} Along the line where Ising-1 critical phase turns into a gapped G1 phase the underlying critical theory at the transition corresponds to the tri-critical Ising superconformal field theory. We use finite-size scaling technique to locate the TCI transition between the Ising-1 and G1 phases. For large-$g$ Ising-2 critical phase turns into G4 gapped phase via yet another TCI line. Due to the presence of incommensurability the location of this critical line is more subtle and we have to look at the extended version of the phase diagram with $h>1$.  The fate of this TCI line inside the Floating-1 phase is not entirely clear.
  \item{\bf Lifshitz line} separates Ising-1 from the Floating-1+Ising phases for $f\lesssim0.18$ when transition between the two is direct. Lifshitz line is a commensurate-incommensurate transition with dynamical critical exponent $z=3$. The critical value of the LL exponent at this transition is $K^c=1$.
  \item{\bf First order transition:} Lifshitz line continues beyond the TCI and disorder lines as a first order transitions separating a region where the parity symmetry is broken for $h<1$ from a region where this symmetry is broken for $h>1$. For a short interval this first order transition quite unusually takes place between the critical Floating-2 and the gapped G3 phases. When the first order transition takes place between G1 and G2 phases, the ground-state along the transition is expected to be 6-fold degenerate. Along the transition between G3 and G4 phases we expect the ground-state to be 12-fold degenerate.
  \item {\bf Kosterlitz-Thouless transitions:}  The Luttinger liquid phase is stable against broken translation symmetry phases if the LL exponent $K>1/p^2$, where $p$ is a periodicity of the symmetry-broken phase. Since Ising-2, G3 and G4 phases all breaks translation symmetry by $p=2$ sites the transition between these three phases and the corresponding floating ones take place when the Luttinger liquid exponent drops to a critical value $K^c=1/4$. On both sides of these transitions the phases are incommensurate that the transitions are of the Kosterlitz-Thouless type\cite{Kosterlitz_Thouless}. On the other side, the LL phase is stable against the pairing instability for $K<1/2$. Note that a pairing - an operator that simultaneously creates or destroys a pair of particles and preserves parity - is very similar to an instability of a spin-flip in the paramagnetic phase since the latter always creates a pair of domain walls. In both cases the operators are relevant and lead to a gapped G1 and G2 phases when the Luttinger liquid parameter exceeds $K^c=1/2$.
    \item{\bf Commensurate line and eight vertex point:} Inside the overall incommensurate G4 and G2 phases there is a line where the wave-vector $q$ takes the commensurate value $q=\pi$. Along this line the floating phase collapses into a direct transition that according to our numerical data belongs to the universality class of the eight-vertex model\cite{BAXTER1972193}. 
 \end{itemize}

\subsection{Duality}


The spin model defined by the Hamiltonian (\ref{eq:spin}) up to boundary terms transforms into itself by Kramers-Wannier duality transformation:
\begin{equation}
  \sigma^x_i\sigma^x_{i+1}\rightarrow\tilde{\sigma}_i^z \ \ \ \mathrm {and} \ \ \   \sigma_i^z \rightarrow \tilde{\sigma}^x_i\tilde{\sigma}^x_{i+1},
\end{equation}
where $\sigma$ and $\tilde{\sigma}$ are Pauli matrices. Coupling constants $f$ and $g$ in the dual model are re-scaled to $g\rightarrow g/h$ and $f\rightarrow f/h$. This implies that in an extended model with $h\neq 1$ the transitions between each pair of the dual phases take place exactly at $h=1$. Therefore a phase diagram presented in Fig.\ref{fig:PD} describes a plane of phase transitions in an extended model of Majorana chain (\ref{eq:majorana}) with alternating hoping $t_\mathrm{even}\neq t_\mathrm{odd}$.

\begin{figure}[t!]
\centering 
\includegraphics[width=0.45\textwidth]{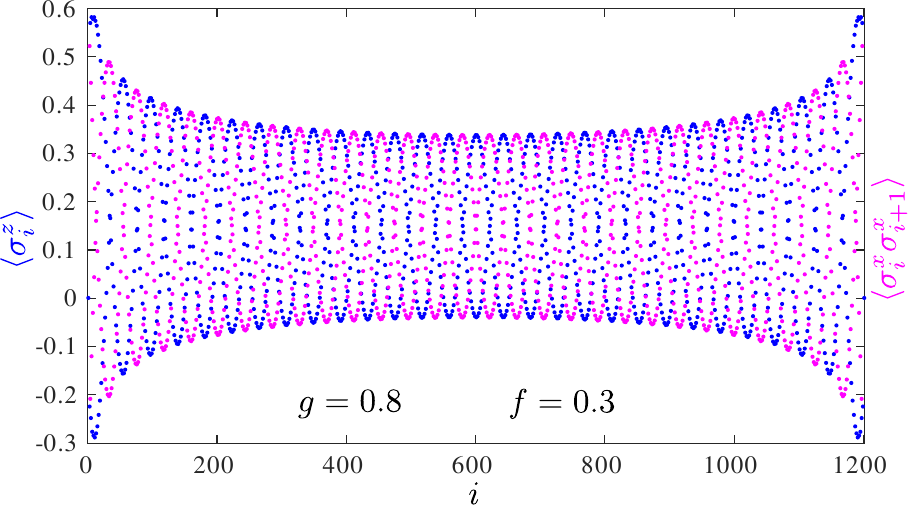}
\caption{{\bf Duality reflected in local observables.} Friedel oscillation profile of $\langle \sigma^z_i\rangle $ (blue) and its dual $\langle \sigma^x_i\sigma^x_{i+1}\rangle$ (magenta) at the boundary of the floating phase at $g=0.8$ and $f=0.3$. Perfect agreement between the envelops and oscillation frequency are the result of self-duality of the model. }
\label{fig:dual}
\end{figure}

The property of duality is also reflected in local observables. As an illustration we show in Fig.\ref{fig:dual}  profiles of local magnetization $\langle \sigma^z_i\rangle$ and its dual $\langle \sigma^x_i\sigma^x_{i+1}\rangle$ that appears on a finite-size chain due to Friedel oscillations.  One can see a perfect agreement between the critical scaling (the envelope of the profile) and between the incommensurate wave-vectors (the frequency of oscillations).

%
%

\section{Tri-critical Ising lines}
\label{sec:TCI1}

\subsection{Location of the tri-critical Ising and disorder lines for small $g$}

In the non-interacting case $f=g=0$ the Hamiltonian (\ref{eq:spin}) reduces to the transverse-field Ising model at the critical point $h=1$.
From the previous studies \cite{rahmani,ObrienFendley} it is known that neither $f$- nor $g$-terms immediately destroy Ising criticality. For $g=0$ it has been shown that upon tuning the coupling constant $f$ the critical Ising phase terminates at $f\approx 0.428$ with the end point in the tri-critical Ising universality class\cite{ObrienFendley}. Beyond the end point the system is gapped with three-fold degenerate ground-states that corresponds to the two states with broken $\mathbb{Z}_2$ symmetry (ferromagnetic along $x$) and one with  $\mathbb{Z}_2$ symmetry preserved (paramagnetic).  At the frustration-free point $f=0.5$ these three states are exact\cite{ObrienFendley}.  Furthermore, this point is also a disorder point beyond which the short-range correlations are incommensurate.

\begin{figure}[t!]
\centering 
\includegraphics[width=0.49\textwidth]{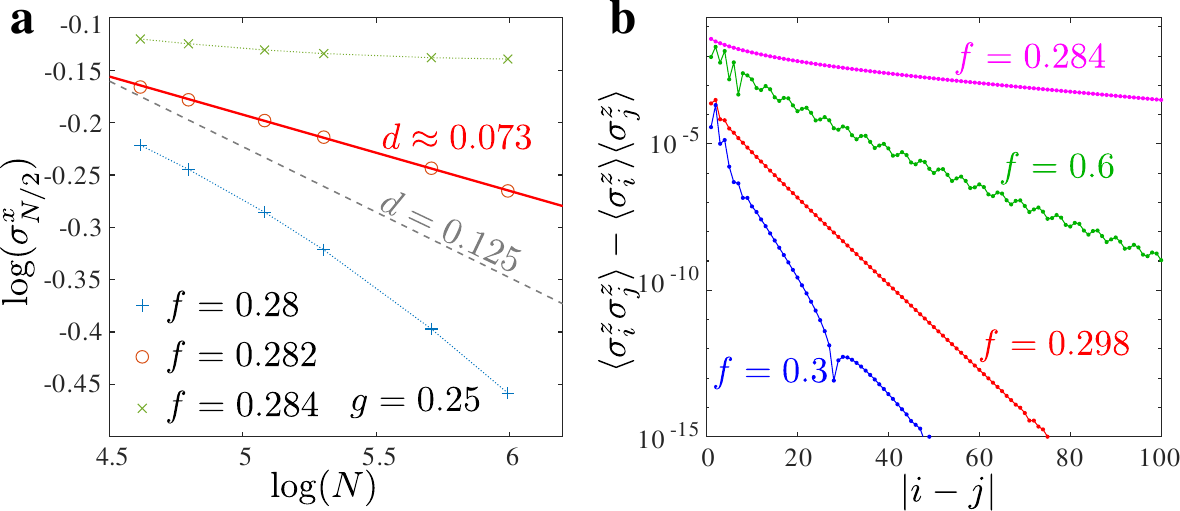}
\caption{{\bf Location of the tri-critical Ising and disorder line for $g=0.25$.} {\bf a} Finite-size scaling of the middle chain magnetization along $x$ with boundary spins polarized in the same direction. Tri-critical Ising point is associated with the separatrix; the slope (red line) is in excellent agreement with the scaling dimension $d=3/40$ of the tri-critical Ising model. The slope with the scaling dimension $d=1/8$ of the Ising critical theory is included for a reference (dashed gray line).  {\bf b} Scaling of the connected correlation function with the distance for $g=0.25$ and various values of $f$. Starting from  $f\approx0.3$ the system shows the presence of incommensurability.}
\label{fig:TCI1}
\end{figure}

For finite $g$ critical properties of the system are qualitatively similar: the Ising critical phase terminates with the tri-critical Ising line beyond which the system enters the gapped phase with a three-fold generate ground-states. Shortly-after the TCI line the system hits disorder line and its short-range order becomes incommensurate. 
 In order to locate the tri-critical Ising line for a finite-$g$ we look at the finite-size scaling of the order parameter. For this we take local magnetization in $x$-direction that should decay algebraically in the critical regime and stays finite in the gapped phase. In order to lift the degeneracy and superposition with the second ferromangetic state we polarized the boundary spins in $x$ direction. This acts as an impurity and lead to a Friedel oscillation profile that according to the boundary conformal field theory scales as $\langle\sigma^x_i\rangle\propto[(N/\pi) \sin (\pi i/N)]^{-d}$, where $d$ is the scaling dimension of the corresponding operator $\sigma$. For the tri-critical Ising minimal model $d=h_\sigma+\bar{h}_\sigma=\frac{3}{80}+\frac{3}{80}=0.075$\cite{difrancesco}. For the Ising critical theory the corresponding scaling dimension is significantly higher  $d=\frac{1}{16}+\frac{1}{16}=0.125$. This allows to identify tri-critical Ising transition with the separatrix in the finite-size scaling of the order parameter in a log-log scale: any convex curve leads to a finite $\sigma^x$ magnetization in the thermodynamic limit and thus to the gapped phase, while the concave curve will eventually get the slop $d=1/8$ of the Ising critical phase. In Fig.\ref{fig:TCI1}(a) we provide an example of such a finite-size scaling for $g=0.25$. Based on the results we locate the tri-critical Ising line at $f\approx0.282$ and the slope of the separatrix $d\approx0.073$ is in excellent agreement with the CFT prediction $d=0.075$.

In order to locate the disorder line away from the exactly solvable point, we look at the connected correlations $\langle\sigma^z_i\sigma^z_{j}\rangle-\langle\sigma^z_i\rangle\langle\sigma^z_{j}\rangle$. In Fig.\ref{fig:TCI1} we provide examples of these correlations function for  $g=0.25$ and various values of $f$ inside the dapped phase. For each point correlations decay exponentially fast with the distance $|i-j|$ but starting from $f\approx0.299$. One can clearly distinguish periodic oscillations with incommensurate wave-vector $q$. The disorder line corresponds to the kink in the correlation length, that is often (though not always) is also a sharp minimum of the correlation length. This agree with a non-monotonous behavior of the slope of the correlations and despite the proximity to the tri-critical Ising point the correlation length at the disorder line is very small (for $f=0.298$ we got $\xi\approx 3.1$). Let us also emphasize that upon tuning the $g$-term the tri-critical Ising and the disorder lines take place at smaller values of $f$ and approach each other.

\subsection{Tri-critical Ising transition for large $g$}

For small values of $g$ we located the tri-critical Ising line by looking as the finite-size scaling at the self-dual plane. Because of the incommensurability at large $g$ this method would require to access much larger systems sizes that would significantly exceed current computational limits. Instead we closely follow the procedure introduced in Ref.\onlinecite{chepiga_laflorencie} - we look at the scaling of the order parameter for $h>1$ as a distance to the self-dual plane $h=1$. 

\begin{figure}[t!]
\centering 
\includegraphics[width=0.38\textwidth]{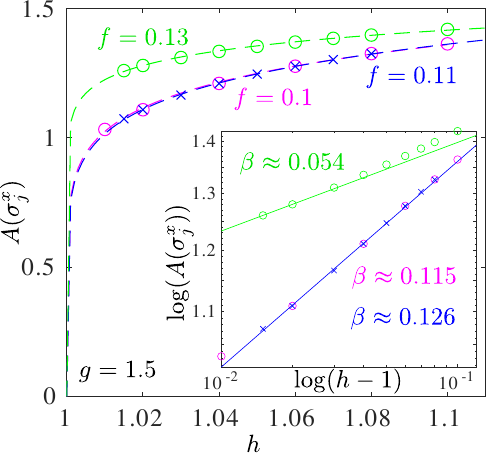}
\caption{{\bf Critical scaling towards Ising and tri-critical Ising transition}. The amplitude $A(\sigma^x_i)$ extracted in the middle of the chain with $N=400$ sites in an extended version of the model with $g=1.5$ and three values of $f=0.1$, 0.11, 0.13 as a function of field $h>1$. Extracted critical exponent $\beta\approx0.115$ and $0.126$ are in excellent agreement with the Ising critical theory, while for $f=0.13$ the critical exponent is significantly smaller in a qualitative agreement with the tri-critical Ising point for which the theory predicts $\beta=1/24$. Inset: the same plot in a log-log scale.}
\label{fig:TCI_2}
\end{figure}

We take an amplitude of the Friedel oscillations $A(\sigma^x_j)$ in the period-2$\mathbb{Z}_2$ phase as an order parameter. The amplitude of the oscillations $A(\sigma^x_j)$ is extracted as a difference between the largest and
smallest values that $\langle\sigma^x_j\rangle$ takes over an interval of length $N/4$ in the middle of the chain. Of course, to see it non-zero we have to break the parity symmetry by polarizing the edge spins in $x$ direction. Upon approaching the Ising transition the amplitude is expected to decay with the Ising critical exponent $\beta=1/8$, while upon approaching the tri-critical Ising point, the critical exponent is much smaller and is equal $\beta=1/24$. Here we locate the tri-critical Ising line following the same procedure. An example for $g=1.5$ is provided in Fig.\ref{fig:TCI_2}. One can see that for $f\leq 0.11$ the scaling is in excellent agreement with the Ising critical theory, while for $f=0.13$ the effective critical exponent is significantly smaller. It is worth to mention that the tri-critical line shown in Fig.\ref{fig:PD} was obtained with $N=400$ sites. We expect that due to finite-size effects associated with the incommensurability the location can be slightly underestimated, thus one can take the blue line in Fig.\ref{fig:PD} as a lower bound of the TCI transition in the thermodynamic limit. As an upper bound, one take the commensurate line.

Upon approaching the Floating-1 phase and the commensurate line the finite-size effects becomes stronger and with the available method we cannot reach the sufficient accuracy to locate the tri-critical Ising line below $g=1.2$. In this respect the fate of the tri-critical Ising line in the middle part of the phase diagram remains an open question. We do not see any indication of the Ising criticality on the right side of the commensurate line. Thus the most feasible scenario is that the tri-critical Ising line makes a turn and together with the second tri-critical Ising transition coming from small-$g$ ends up at the end point of the Lifshitz line. We will come back to this question in Section \ref{sec:floating} while discussing the floating phases.

%

\section{Lifshitz transition}
\label{sec:Lifshitz}

\begin{figure}[t!]
\centering 
\includegraphics[width=0.49\textwidth]{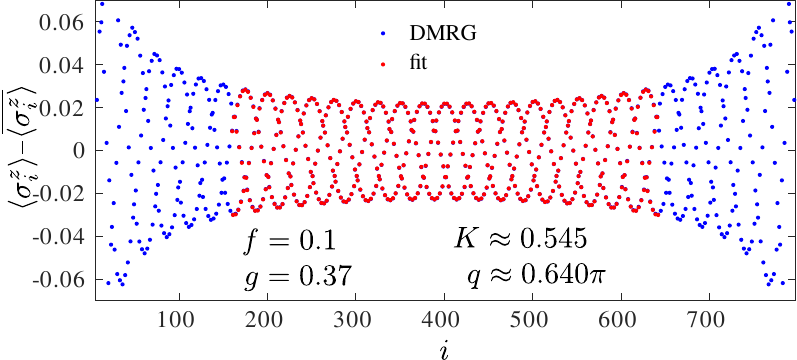}
\caption{{\bf Extraction of floating phase characteristics from Friedel oscillations profile.} Examples of the Friedel oscillations inside the floating phase obtained on a finite-size system with $N=802$ sites with polarized boundary conditions. Blue points are DMRG data, red points are the result of the fit with Eq.\ref{eq:friedel} (blue dots are completely hidden under red ones). Note that the uniform part of the spin density ${\overline{\langle \sigma^z_i\rangle}}$ has been subtracted, and the fitting window is restricted to the range $i \in[160, 642]$ in order to avoid a short-distance corrections.}
\label{fig:float}
\end{figure}

 Upon tuning the coupling constant $g$ and for small enough $f$ the system undergoes a Lifshitz transition and enters the floating phase. 
At every point floating phases can be characterized by the Luttinger liquid exponent $K$ and the incommensurate wave-vector $q$. We extract both quantities by fitting the Friedel oscillations profile with\cite{3boson}
 \begin{equation}
    \langle\sigma^z_j\rangle \propto \frac{\cos(q j)}{[(N/\pi) \sin (\pi j/N)]^{K}}.
    \label{eq:friedel}
 \end{equation}
In Fig.\ref{fig:float} we show an example of such a fit for $f=0.1$ and $g=0.37$. 
In order to reduce the edge effects, we discard $20\%$ of sites close to each end of the chain and only fit the middle part part as shown in Fig.\ref{fig:float}. The result of the fit (red) is in an excellent agreement with DMRG data (blue), such that the latter are completely hidden under the red dots. This method allows us to extract $K$ and $q$ with the spectacular accuracy.

\begin{figure}[t!]
\centering 
\includegraphics[width=0.49\textwidth]{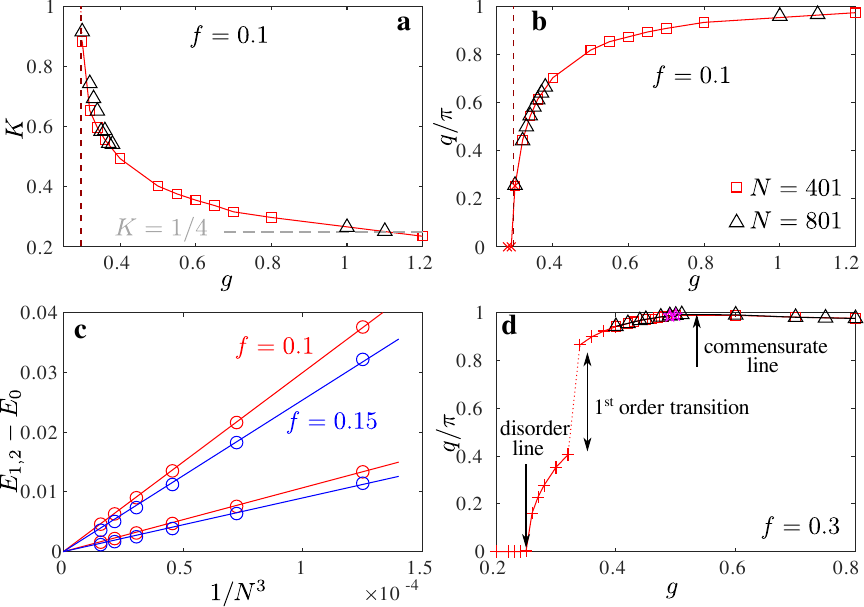}
\caption{{\bf Numerical evidences of the Lifshitz critical line.} {\bf a} Luttinger liquid exponent $K$ and {\bf b} wave-vector q as a function of coupling $g$ along a vertical cut at $f=0.1$. $K$ takes the critical value $K^c=1$ at the Lifshitz transition at $g\approx0.295$ (dashed dark-red line), at the same point the wave-vector $q$ shown in {\bf b} develops incommensurability. The Luttinger liquid phase terminates at $K^c=1/4$. {\bf c} Finite-size scaling of the energy gap between the ground-state ($E_0$) and first ($E_1$) and second ($E_2$) excitations for $f=0.1$(red) and 0.15 (blue) as a function of $N^{-3}$. Linear scaling is in the perfect agreement with the dynamical critical exponent $z=3$ of the Lifshitz transition. {\bf d} Wave-vector $q$ along a vertical cut at $f=0.3$. In the gapped phases the wave-vector is extracted by fitting short-range correlations (red crosses); in the critical phases - by fitting the Friedel oscillation profiles (red crosses for $N=401$, black triangles for $N=801$). There is a pronounced jump in the wave-vector $q$ at the first order transition that appears as a continuation of the Lifshitz critical line.}
\label{fig:lifshitz}
\end{figure}

 Along $f=0$ line the transition to the floating phase is known to be of the Lifshitz type\cite{rahmani, chepiga_laflorencie}.  Lifshitz transition is a very special critical point at which (in addition to the Ising criticality) the system simultaneously enters the  Luttinger liquid phase and develops incommensurability. We check this by looking at the LL exponent $K$ and the wave-vector $q$ as a function of coupling $f$ as shown in  Fig.\ref{fig:lifshitz}{\bf a-b}. The Luttinger liquid is destroyed and the system undergoes a Lifshitz transition at $g\approx0.295\pm0.005$ when the LL exponent reaches it critical value $K^c=1$\cite{chepiga_laflorencie}. At the same value of $g$ the wave-vector $q$ shown in Fig.\ref{fig:lifshitz}{\bf b} starts to be incommensurate. 
Note that the finite-size effects at this transition are negligibly small.

Above we have introduced the Lifshitz transition  as a critical point where Luttinger liquid phase emerges together with incommensurability. In the theory of quantum phase transitions there is a second example that fits this definition - the Pokrovsky-Talapov transition\cite{Pokrovsky_Talapov}. Both transitions are characterized by the vanishing sound velocity that leads to a dynamical critical exponent $z>1$, thus none of the two are conformal. But the dispersion relations and thus the exact values of the dynamical exponents $z$ are different: for Pokrovsky-Talapov transition $z=2$, while for the Lifshitz universality class the theory predicts $z=3$.  In order to check that the transition that we face in the phase diagram of Fig.\ref{fig:PD} is indeed the Lifshitz one we look at the finite-size scaling of the energy gap that is expected to vanish as  $\Delta\propto N^{-z}$. We extract low-lying levels of the excitation spectra for systems with up to 40 sites  by targeting several states along with the ground-state in DMRG simulations as described in Ref.\onlinecite{dmrg_chepiga}. The results of these simulations for $f=0.1$ and $0.15$ are presented in Fig.\ref{fig:lifshitz}(c) and are in excellent agreement with $z=3$.

Another distinct feature of the Lifshitz transition is that by contrast to the Pokrovsky-Talapov one it appears as a multi-critical point, or in the present case - a multi-critical line. In order to see that one has to think in terms of an extended model, for instance, the one  with $h\neq 1$, and look at the phases away from the self-dual plane. This extended model has been explored recently for $f=0$\cite{chepiga_laflorencie}: there, the Lifshitz point located at  $g^c\approx0.29$ and $h^c=1$ appears as a multicritical point of four phases: the $\mathbb{Z}_2$-symmetry broken phase for $h<h^c$ and $g<g^c$; its dual - the paramagnetic phase at $h>h^c$ and $g<g^c$; the ordinary floating phase for $h<h^c$ and $g>g^c$; and a dual to that - the floating phase with spontaneously broken $\mathbb{Z}_2$ symmetry for $h>h^c$ and $g>g^c$. In other words, in the phase diagram shown in  Fig.\ref{fig:PD} the $\mathbb{Z}_2$ symmetry is broken for $h<1$ below the Lifshitz line and for $h>1$ - above it.

\subsection{First order transition}

\begin{figure}[t!]
\centering 
\includegraphics[width=0.49\textwidth]{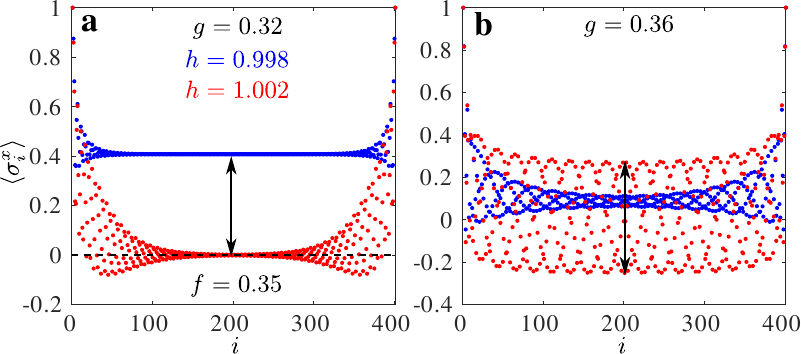}
\caption{{\bf Location of the $\mathbb{Z}_2$ broken symmetry phase on two sides of the first-order transition}.  Friedel oscillations from boundary spins polarized in $x$-direction for $f=0.35$ and {\bf a} $g=0.32$ below the first order transition and {\bf b} $g=0.36$ above it. For each point we present two profiles away from the self-dual surface: for $h=1.002$ (red) and for $h=0.998$ (blue). The order parameter that reflects broken $\mathbb{Z}_2$ symmetry is indicated with black arrows (for details see the main text).}
\label{fig:two_points}
\end{figure}

Surprisingly, after the Lifshitz line meets the tri-critical Ising transition it continues as a first order transition. At this transition we observe a pronounced jump in the wave-vector $q$ as shown in Fig.\ref{fig:lifshitz}{\bf d}. 
 Below the first order transition the $\mathbb{Z}_2$-broken symmetry phase  is realized for $h<1$, while above the first order line - the $\mathbb{Z}_2$ symmetry is broken for $h>1$.
 We illustrate this in Fig.\ref{fig:two_points} where we present Friedel oscillations from the edge spins polarized in $x$-direction for two points - below and above the first order line and away from the self-dual plane. Below the first order transition at $f=0.35$ and $g=0.32$ and for $h<1$ the ground-state breaks $\mathbb{Z}_2$ symmetry and corresponds to the two ferromagnetic states polarized along $x$. Imposed boundary conditions picks up one of these two states and allow us to detect a finite magnetization $\langle\sigma_i^x\rangle$ in the bulk. At the same values of couplings $f$ and $g$ but for $h>1$ we see that $\sigma^x_i$ is zero in the bulk that is consistent with unbroken $\mathbb{Z}_2$ symmetry. For the point above the first order line, $f=0.35$ and $g=0.36$, we observe an opposite situation. Following Ref.\onlinecite{chepiga_laflorencie} we associate a $\mathbb{Z}_2$ order parameter with an amplitude of the oscillations: it is finite for $h>1$ and goes to zero (although quite slowly due to proximity to the floating phase) for $h<1$.
 
\section{Floating phases}
\label{sec:floating}

Let us now take a closer look at the floating phases and their boundaries. 
 The Luttinger liquid phase is stable against broken translation symmetry by two sites if the LL exponent exceeds the critical value $K^c=1/4$. In all three phases in the upper part of the phase diagram - Ising-2, G3 and G4 - the translation symmetry is broken (in fact, the translation symmetry is broken in the same fashion even away from the self-dual plane\cite{chepiga_laflorencie}. This implies that the transition out of all floating phases - Floating-1, 2 and 3 - will take place at the same value of the LL exponent $K^c=1/4$.

\begin{figure}[t!]
\centering 
\includegraphics[width=0.5\textwidth]{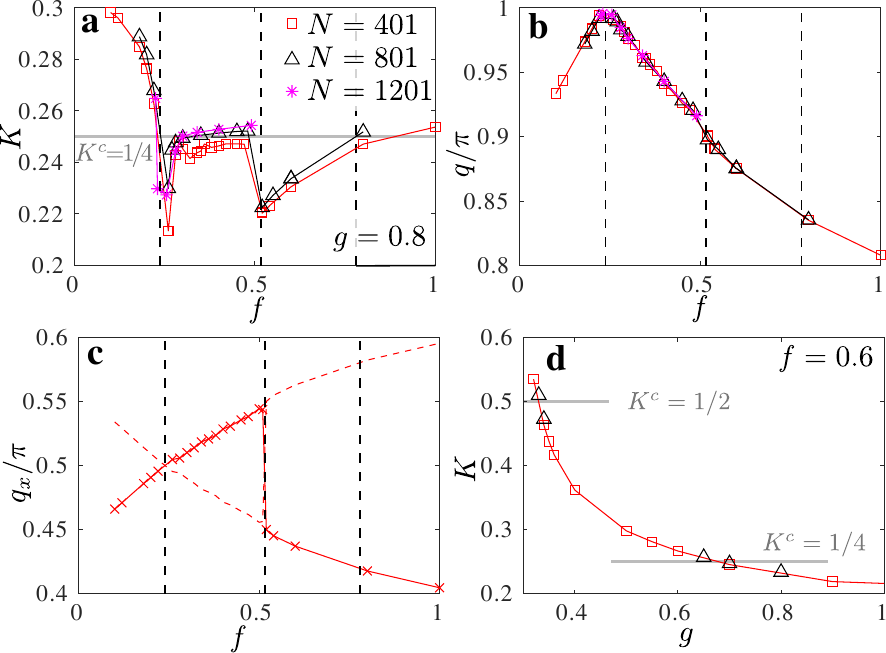}
\caption{{\bf Numerical results for the floating phases}. {\bf a} Luttinger liquid exponent $K$, {\bf b} wave-vector  $q$ of   $\sigma^z$ and {\bf c} wave-vector $q_x$ of $\sigma^x$ correlations. In {\bf a} one can see the presence of four quantum phase transitions. $K$ drops below $K^c=1/4$ on both sides of the commensurate line where $q=\pi$ and $q_x=\pi/2$ implying that there is no floating phase along the commensurate line, thus there are two Kosterlitz-Thouless transitions on both sides of it. Around $f\approx0.5$ the system undergoes a first order transition from the Floating-2 to G3 phase: $K$ shows a finite jump;  $q_x$ (red crosses) jumps to $1-q_x$ (dashed line). Around $f\approx0.8$ the system undergoes yet another Kosterlitz-Thouless transition between G3 and Floating-3 phases.  {\bf d} Luttinger liquid exponent along a vertical cut at $f=0.6$. The Floating-3 is stable when the LL exponent $1/4<K<1/2$.}
\label{fig:floating2}
\end{figure}

 An example of the Luttinger liquid parameter $K$ extracted along a horizontal cut at $g=0.8$ that goes through all three floating phases and phases G3 and G4 is shown in Fig.\ref{fig:floating2}{\bf a}. One can see that $K$ is not monotonous and has two pronounced drops. The first one takes place around $f\approx0.25$ where the wave-vector $q$ takes commensurate value $q=\pi$ (see Fig.\ref{fig:floating2}{\bf b}). This wave-vector measures incommensurability in local magnetization $\langle\sigma^z_i\rangle$, and, because of the duality, in the pairing $\langle\sigma_i^x\sigma_{i+1}^x\rangle$. In Fig.\ref{fig:floating2}{\bf c} we present the incommensurate wave-vector $q_x$ computed for $\sigma^x_i$ operator. One can see that around $f\approx 0.25$ it continuously passes the value $q_x=\pi/2$. The floating phase cannot exists along this commensurate line and thus collapses into a single transition point that we will discuss in details in the next section. The fact that on both sides of the commensurate line the Luttinger liquid exponent drops below $K^c=1/4$ fully agrees with this picture. The second drop takes place around $f\approx0.5$ and is associated with the first order transition from the Floating-2 phase to the G3 phase until eventually, around $f\approx0.78$ the Luttinger liquid parameter $K$ exceeds the critical value $K^c=1/4$ and the system enters the Floating-3 phase. 

In the lower part of the phase diagram in Fig.\ref{fig:PD} the Luttinger liquids become unstable due to  pairing instability that becomes relevant when the Luttinger liquid parameter exceeds $K^c=1/2$.
Since G1 phase is always incommensurate in the vicinity of the floating phases the transition is of the Kosterlitz-Thouless type. In  Fig.\ref{fig:floating2}{\bf d} we provide an example of the Luttinger liquid exponent $K$ along the vertical cut at $f=0.6$. We associate the boundaries of the Floating-3 with the two points where the LL exponent $K$ takes critical values $K^c=1/2$ and $1/4$. Similar procedure has been applied for a transition  between the G1 Floating-1 phases. We also expect that the transition between the floating-2 and G1 phases takes place when $K=1/2$ but we cannot confirm this numerically due to proximity of the first order transition between G1 and G2 phases and the commensurate line.

 Let us also comment here that the pairing instability responsible for the transitions into G1 and G2 phases are not relevant in the critical Ising phase.  Therefore the critical value of the Luttinger liquid exponent at the Lifshitz transition is larger and equal to $K^c=1$\cite{chepiga_laflorencie}.

\begin{figure}[t!]
\centering 
\includegraphics[width=0.4\textwidth]{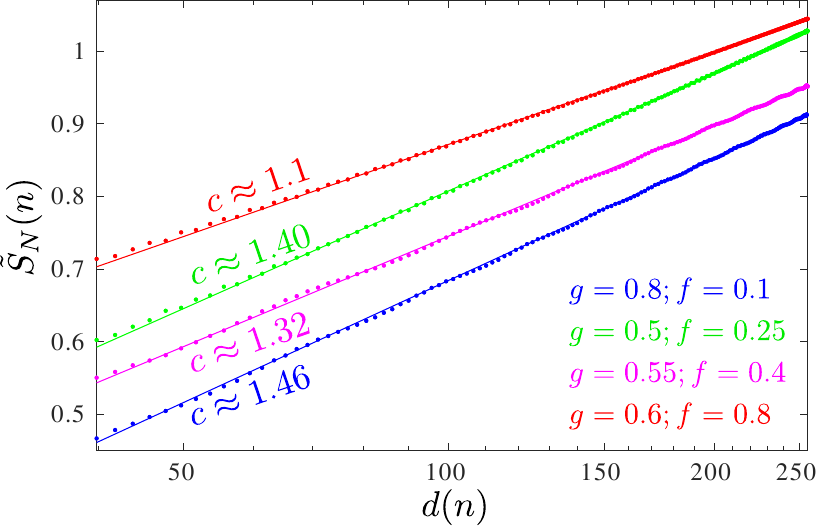}
\caption{{\bf Finite-size scaling of the reduced entanglement entropy $\tilde{S}_N(n)$ with the conformal distance $d(n)$}. The central charges indicated for each curve were obtained by fitting the $\tilde{S}_N(n)$ with Eq.(\ref{eq:calabrese_cardy}). Except the points deep inside the Floating-1+Ising phase (blue) the central charge always significantly smaller that $c=3/2$}
\label{fig:cc}
\end{figure}

In order to distinguish pure floating phases from the critical phase whe the Luttinger liquid is superposed with Ising criticality we extract the central charge.
In conformal field theory\cite{difrancesco} Luttinger liquid phase is characterized by the central charge $c=1$, while Ising criticality has the central charge $c=1/2$.
When the two critical regimes come together the phase is characterized by the central charge $c=1+1/2=3/2$. The entanglement entropy $S_N(n)=-\mathrm{Tr} \rho_n \ln \rho_n$ is extracted from the eigenvalues of the reduced density matrix $\rho_n$.
We extract the central charge numerically from the finite-size scaling of the entanglement entropy in an open chain that on a finite chain with $N$ sites scales with the size of the subsystem $n$ ($1\ll n\ll N$) as\cite{CalabreseCardy}:
\begin{equation}
\tilde{S}_N(n)=\frac{c}{6}\ln d(n)+\zeta \langle {\sigma^z}_n{\sigma^z}_{n+1} \rangle+s_\mathrm{B},
\label{eq:calabrese_cardy}
\end{equation}
where $d=\frac{2N}{\pi}\sin\left(\frac{\pi n}{N}\right)$ is the conformal distance  and $\zeta$ is a non-universal constant introduced in order to suppress Friedel oscillations from fixed boundary conditions\cite{PhysRevLett.96.100603,capponi}, $s_\mathrm{B}$ is a non-universal constant that includes, in particular, the boundary entropy.
 In Fig.\ref{fig:cc} we provide examples of the scaling of the entanglement entropy at four points in the floating phases: Deep inside the Floating-1+Ising phase at $g=0.8$ and $f=0.1$ (blue) the central charge $c\approx1.46$ agree within $3\%$ with $c=3/2$. But by upon approaching the tip of the Floating-1 phase the central charge is systematically smaller. For instance, at $g=0.5$ and $f=0.25$ it takes a value $c\approx1.4$. From previous experience with central charges inside and in the vicinity of the Floating+Ising phase we know that it changes extremely slowly\cite{chepiga_laflorencie}. In this respect the value $c\approx1.4$ is a smoking gun suggesting that at the tip of the floating phase the central charge in the thermodynamic limit can actually drop to $c=1$ but because of huge finite size effect we observe much larger value. If that is the case, then Ising criticality terminates inside the Floating-1 phase.  In the Floating-2 ($g=0.55$, $f=0.4$) and Floating-3 ($g=0.6$, $f=0.8$) phases the central charge is always significantly smaller that $c=3/2$, thus we conclude that Ising criticality does not intervene these floating phases.

\section{The 8-vertex multi-critical point}
\label{sec:thetpoint}

\begin{figure}[t!]
\centering 
\includegraphics[width=0.35\textwidth]{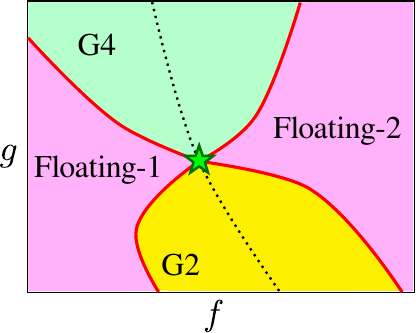}
\caption{{\bf Sketch of the phase diagram around the multi-critical point.} Along the commensurate line (dotted line) the floating phases collapse  and the transition between gapped G2 and G4 phases is direct in the 8-vertex universality class. Red lines state for the Kosterlitz-Thouless transitions}
\label{fig:sketch}
\end{figure}

In Fig.\ref{fig:lifshitz}{\bf d} have already seen that incommensurate wave-vector $q$ eventually takes a commensurate value $q=\pi$.
 Along the commensurate line the floating phase cannot exist and the transition between the G4 and G1 phases has to be direct as sketched in Fig.\ref{fig:sketch}. Recently the nature of this transition has been studied in details in closely related models of interacting Kitaev chain\cite{chepiga8vertex}. In the simplest integrable case the model can be mapped to an XYZ model with $J_x=-J_z$ for which Baxter has shown\cite{BAXTER1972193} that in the vicinity of the transition the critical behavior is governed by the universality class of the eight-vertex model. This implies that critical exponents, although not fixed to a single universal value, all depend on a single parameter $\mu$. In particular, the critical exponent of the order parameter is given by $\beta=(\pi-\mu)/(4\mu)$ and the scaling dimension - the ratio between $\beta$ and the correlation length critical exponent $\nu$ - is $d=\beta/\nu=(\pi-\mu)/(2\pi)$. Later it was numerically established that even in the non-integrable case the direct transition between period-2 and $\mathbb{Z}_2$ phases along a commensurate line belongs to the eight-vertex universality class\cite{chepiga8vertex,chepiga_laflorencie}. In all these cases the location of the commensurate line was known exactly due to an explicit particle-hole symmetry of the Hamiltonian (or equivalently a spin-flip symmetry $\sigma^z_i\rightarrow-\sigma^z_i$).

In the present case, the model defined in Eq.(\ref{eq:spin}) is not invariant under spin-flip and the exact location of the commensurate line is unknown.  Therefore, as a first step we locate the commensurate line by extracting the wave-vector $q$ and identifying the point where the it is commensurate ($q=\pi$) as shown in Fig.\ref{fig:floating2}{\bf b}. We then repeat this procedure along many horizontal cuts. 
In the Appendix \ref{sec:comline} we provide a table with obtained data points. 

\begin{figure}[t!]
\centering 
\includegraphics[width=0.49\textwidth]{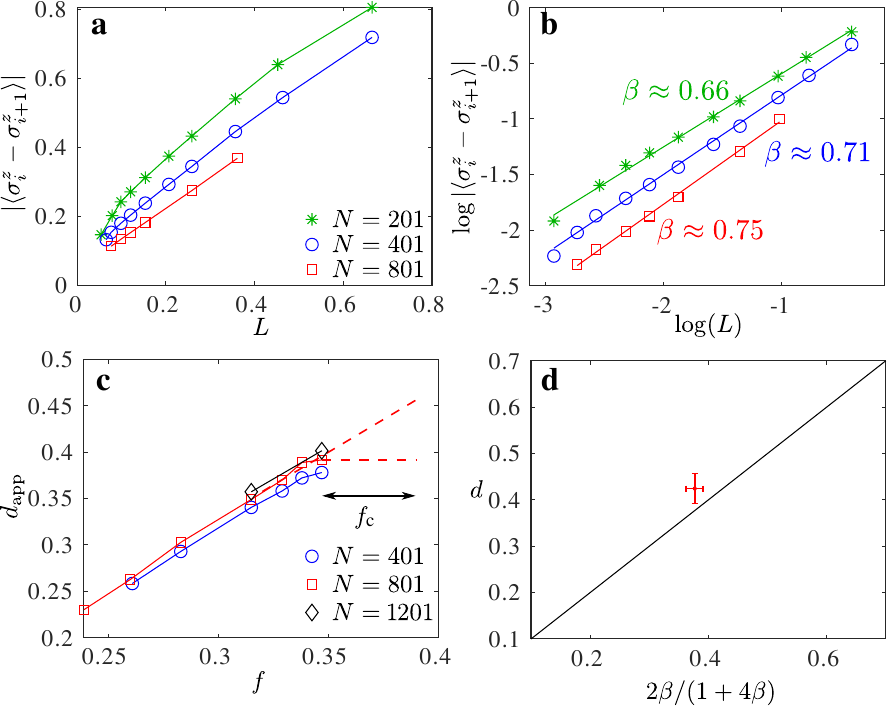}
\caption{{\bf Numerical evidences of the multi-critical point in the eight-vertex universality class.} {\bf a} Local order parameter of the G4 phase with broken translation symmetry $|\langle\sigma_i^z-\sigma_{i+1}^z\rangle|$ as a function of distance $L$ to the transition at $f\approx0.39$. {\bf b} Same as panel {\bf a} but in a log-log scale. The slope of the scaling gives critical exponent $\beta\approx0.66$, 0.71 and 0.75 for chains with $N=201$, 401 and 801 sites. {\bf c} Apparent scaling dimension $d_\mathrm{app}$ as a function of $f$ along the commensurate line. Upper and lower bound of the scaling dimension at the multicritical point are marked with dashed lines. {\bf d} Comparison of the numerically extracted critical exponents $\beta$ and $d$ (red error bars) and the theory predictions for the eight-vertex model (black line).}
\label{fig:thepoint}
\end{figure}

Next, along the commensurate line we measure the local order parameter of the G4 phase. 
Note that, in principle, it is not necessary to follow the commensurate line and the nature of the transition can be extracted along any cut that lies inside the G2 and G4 phases and goes through the multi-critical point. However, in practice, the G4 phase in the vicinity of the multicritical point is extremely narrow, while the location of the multicritical point is unknown, thus the simplest choice for the cut is to follow the commensurate line.
Since all ground states in the G4 phase break translation symmetry we  define the local order parameter as an amplitude of the Friedel oscillations $|\langle\sigma_i^z-\sigma_{i+1}^z\rangle|$ in the middle of the chain. In order to reduce finite-size corrections we fix boundary spins to be polarized along $z$.  
Usually one extracts critical exponent $\beta$ by looking how the order parameter vanishes with the distance $j-j^c$ to the transition, where $j$ is a single tuning parameter that drives the system through the transition located at $j^c$. According to conformal field theory $|\langle\sigma_i^z-\sigma_{i+1}^z\rangle|\propto (j-j^c)^\beta$. And now we face yet another difficulty: the commensurate line along which we have to locate the transition is not a linear function of coupling constants $f$ and $g$. Thus the distance to the transition point has to be computed along the commensurate line. The simplest way to do so is by summing up all intervals between the available points, given the high density of data points this provides a reasonable approximation for $L$. In Fig.\ref{fig:thepoint}{\bf a} we plot an order parameter $|\langle\sigma_i^z-\sigma_{i+1}^z\rangle|$ for three different system sizes $N$ as a function of distance $L$. Here the origin $L=0$ is associated with $f\approx0.39$ and $g\approx0.368$. In the Appendix \ref{sec:more8vert} we also show how the order parameter as a function of $f$ and $g$ along the commensurate line.
Important to notice that all these plots are consistent with continuous transition. 

In order to extract the value of the critical exponent $\beta$ we plot the obtained order parameter as a function of distance to the transition in a log-log scale. Since the location of the critical point is not known we try several possible locations along the commensurate line. The best agreement with the linear scaling is achieved when the critical point is located at $f_c\approx 0.39$, these results are presented in Fig.\ref{fig:thepoint}. In this case the critical exponent for $N=201$ is equal to $\beta\approx 0.66$, the one for $N=401$ is $\beta\approx0.71$ and for $N=801$  it is $\beta\approx0.75$.  By varying the location of the critical point within the interval $0.38\lesssim f\lesssim 0.4$ we got a reasonable agreement with the linear scaling and critical exponent in the range $0.65\lesssim \beta\lesssim 0.9$.  Beyond this interval the scaling clearly deviates from the linear decay. The obtained range of the critical exponent might seems large at the first glace, but it is important to remember that at the eight-vertex critical point $\beta$ can take any value between $0$ and $\infty$. In this respect the interval $0.65\lesssim \beta\lesssim 0.9$ is rather well defined. Let us also comment here that from the previous experience with eight-vertex criticality we know that the size of the period-2 phase is typically slightly overestimated. This means that $f\approx 0.39$ defines an upper bound of the location of the critical point.

\begin{figure}[t!]
\centering 
\includegraphics[width=0.49\textwidth]{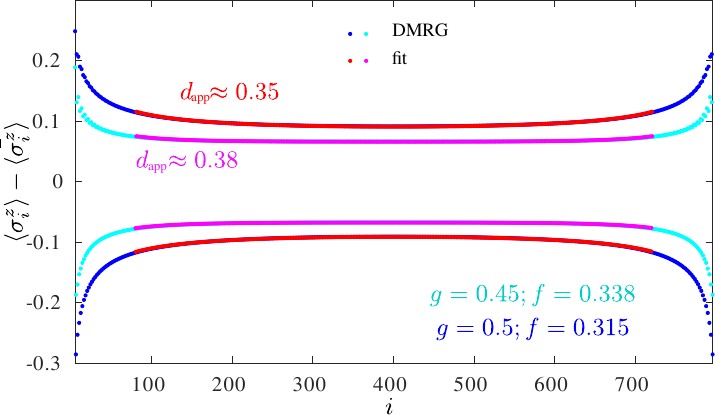}
\caption{{\bf Friedel oscillation profiles along the commensurate line} The profiles are computed for a finite-size chain with $N=801$ sites and boundary spins polarized along $z$-direction. Magenta and red dots are the results of the fit to $\sigma^z_i\propto\cos(\pi i)[(N/\pi) \sin (\pi i/N)]^{-d}$ }
\label{fig:cfriedel}
\end{figure}

We compliment our results for $\beta$ by extracting the scaling dimension $d=\beta/\nu$. At the quantum critical point it can be extracted by fitting the Friedel oscillation profile with $\sigma^z_i\propto\cos(\pi i)[(N/\pi) \sin (\pi i/N)]^{-d}$. Since the exact location of the critical point is not known we will extract the apparent scaling dimension $d_\mathrm{app}$ along the commensurate line. An example of the Friedel oscillation profile along the commensurate line and the fit to the CFT predition with an apparent scaling dimension $d_\mathrm{app}$ is shown in Fig.\ref{fig:cfriedel}.

The results for $d_\mathrm{app}$ are summarized in Fig.\ref{fig:thepoint}{\bf c}. Beyond $f\approx0.35$ the fit is no longer good  and we cannot extract the scaling dimension accurately. However, $d_\mathrm{app}$ seems to be almost linear as a function of $f$. We thus estimate the upper bound of scaling dimension at the critical point $d$ by extrapolating the last four available points to the approximate location of the phase transition $f\approx0.39$. As a lower bound we take the scaling dimension at the last available point where the fit is still good.

To summarize, we end up with the following intervals: $0.65\lesssim \beta \lesssim 0.9$; $0.392\lesssim d \lesssim 0.457$ and the location of the critical point between $0.35\lesssim f\lesssim 0.39$. In order to check whether the transition along the commensurate line indeed belongs to the eight-vertex universality class 
we compare the relation between $\beta$ and $d$ with the theory predictions. Since the  parameter $\mu$ introduced by Baxter\cite{BAXTER1972193} is not known for the non-integrable model, we exclude $\mu$ from the two equations and express $d$ as a function of $\beta$. We end up with the following theory prediction:  $d=2\beta/(1+4\beta)$. In Fig.\ref{fig:thepoint}{\bf d} we show how numerically extracted critical exponents agree with this theory prediction. Given the number of obstacles we had to overcome and the fact that there is no fitting parameter and the comparison between the theory and numerics is direct, the agreement between the two is spectacular. However, let us put as a disclaimer that the shown error bars mainly account for the error associated with the location of the critical point on a commensurate line. It is important to bare in mind that there is also an error that comes from our estimate of distances to the transition point along the commensurate line, from finite size effects, etc. But even keeping all these in mind there is a little room for doubts about the nature of this multicritical point. 

\section{Discussion}
\label{sec:discussion}

 To summarize, in the present paper we study the ground-state properties of the interacting Majorana chain. We have shown that the combination of the two interaction terms of the shortest possible range lead to an extremely rich phase diagram and a wide variety of critical phenomena. We hope this will motivate further exploration of frustrated Majorana chains.
 
 The most striking result reported in the paper is an emergence of the commensurate line along which the floating phase collapses into a single multicritical point. We have provided numerical evidences that this point belongs to the universality class of the eight-vertex model. Thus far, the eight vertex universality class has only been realized along the particle-hole (or spin-flip) symmetry lines. However, the Hamiltonian (\ref{eq:spin}) does not preserve this symmetry explicitly. It means that along the commensurate line this symmetry must be emergent.
 
  For $f=0$ the eight vertex critical point has been realized at $h=0$ between period-2 and $\mathbb{Z}_2$ phases. Based on the duality, it was argued that the same critical point must appear at $h=\infty$ between the paramagnetic and period-2-$\mathbb{Z}_2$ phase. According to our numerical results, by tuning coupling constant $f$ one can bring these two points to a self-dual surface and realized the two copies of the eight vertex model simultaneously. It implies that by looking at the extended phase diagram with $h\neq 1$ one should be able to  track the two multi-critical points on their way towards each other. It also implies that in a 3D parameter space of $(h,g,f)$ there is a 2D surface  where correlations are commensurate. 
 
 Another interesting feature is an existence of an extended Lifshitz line with the dynamical critical exponent $z=3$. Typically appearing as a multicritical point, Lifshitz criticality extends here over a finite interval thanks to the self-duality of the model and it is indeed a line of multicritical points in an extended version of the model with three-dimensional parameter space.
  Interesting to notice that at least one (and probably two) tri-critical Ising lines meet the Lifshitz transition at its end point. The nature of this end point remains an open question and is left for future investigation.
  
The tri-critical Ising conformal field theory is supersymmetric\cite{difrancesco,ObrienFendley}, thus we might expect the supersymmetry to emerge along both tri-critical Ising lines. In addition to that, the entire Floating-1+Ising phase might have an emergent $\mathcal{N}=(1,1)$ supersymmetry\cite{FODA1988611,PhysRevLett.102.176404}. The conditions to that are preserved  $\mathbb{Z}_2$ and U(1) symmetries. The former is preserved by the model and not spontaneously broken at the Ising transition at the self-dual plane. The latter is an emergent symmetry that stabilizes the floating phase\cite{verresen,rahmani}. Furthermore, along the Kosterlitz-Thouless transition between the Floating-1+Ising and Ising-2 critical phase one might expect the spontaneously emergent $\mathcal{N}=(3,3)$ supersymmetry. According to Ref.~\onlinecite{huijse_mult} this higher supersymmetry can be realized if  the velocity of the fermionic degree of freedom is smaller than or equal to the velocity of the bosonic degree of freedom. The verification of this condition goes beyond the scope of the present paper and is left for future studies. Finally, there is also a line where supersymmetric tri-critical Ising line is superposed with the floating phase and point where it crosses the Kosterlitz-Thouless transition and enters the critical phase. It would be very interesting to understand the underlying critical theory and the type of supersymmetries  emergent in these two cases.


\section{Acknowledgments}

I am indebted to Nicolas Laflorencie and Erez Berg for an insightful  discussions.
 This research has been supported by the Delft Technology Fellowship. Numerical simulations have been performed at the Dutch national e-infrastructure with the support of the SURF Cooperative and at the DelftBlue HPC.

\begin{appendix}
  
\section{The location of the commensurate line}
\label{sec:comline}

In this section we provide additional numerical data along the commensurate line.
In the Table below we list the location of the available data points along the commensurate line and the distance to the multi-critical point.

\begin{center}
\begin{table}
  \begin{tabular}{ l |  c |  r }
  $g$ & $f$ & $L(f\approx 0.39)$ \\
  \hline
  1 & 0.214 & 0.6649 \\
  0.8 & 0.241 & 0.4521 \\
  0.7 & 0.261 & 0.3565 \\
  0.6 & 0.283 & 0.2585 \\
  0.55 & 0.297 & 0.2065 \\
  0.5 & 0.315 & 0.1534 \\
  0.47 & 0.329 & 0.1202 \\
  0.45 & 0.338 & 0.0983 \\
  0.43 & 0.347 & 0.0788 \\
  0.41 & 0.358 & 0.0535 \\
  0.4 & 0.364 & 0.0419 \\
  0.38 & 0.38 & 0.0162 \\  
  0.36 & 0.397 & \\
\end{tabular}
\caption{The table lists the location of the available data points along the commensurate line. In the last column we also provide our estimate of the distance along the commensurate line to the possible location of the critical point at $f\approx0.39$.}
\label{tab:table}
\end{table}
\end{center}

\section{The order parameter along the commensurate line}
\label{sec:more8vert}

\begin{figure}[t!]
\centering 
\includegraphics[width=0.49\textwidth]{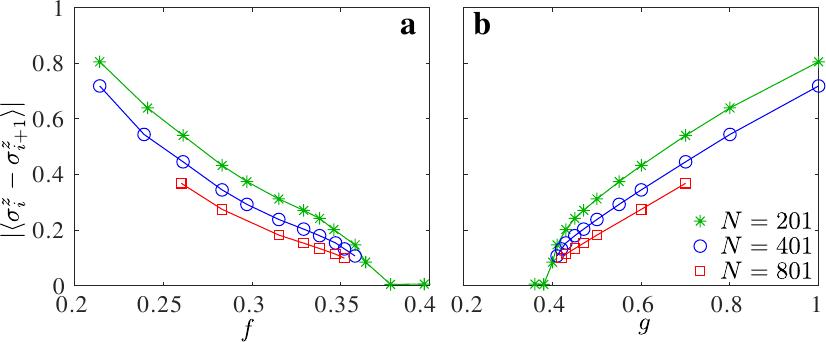}
\caption{{\bf Local order parameter of the G4 phase}. We show the order parameter $|\langle\sigma_i^z-\sigma_{i+1}^z\rangle|$ as a function {\bf a} $f$ and {\bf b} $g$ along the commensurate line. In both cases the results are consistent with continuous transition}
\label{fig:eightpoint2}
\end{figure}

In Fig.4{\bf a} of the main text we have shown how the order parameter scales with the distance $L$ to the transition. In Fig.\ref{fig:eightpoint2} we show similar scaling as a function of coupling constants $f$ and $g$. In both cases the results are consistent with continuous transitions.

\end{appendix}

\bibliographystyle{ieeetr}
\bibliography{bibliography}

\end{document}